\documentclass[aps,prb,twocolumn,groupedaddress,showpacs,superscriptaddress]{revtex4}

\usepackage{graphics,graphicx}
\usepackage{enumitem}

\begin{document}

\title{Homogenization of resonant chiral metamaterials}

\author{Andrei Andryieuski}
\email[]{andra@fotonik.dtu.dk} \affiliation{DTU Fotonik - Department
of Photonics Engineering, Technical University of Denmark,
{\O}rsteds pl. 343, DK-2800 Kongens Lyngby, Denmark}
\author{Christoph Menzel}
\affiliation{Institute of Condensed Matter Theory and Solid State
Optics, Friedrich-Schiller-Universit\"at Jena, Max-Wien-Platz 1,
D-07743 Jena, Germany}
\author{Carsten Rockstuhl}
\affiliation{Institute of Condensed Matter Theory and Solid State
Optics, Friedrich-Schiller-Universit\"at Jena, Max-Wien-Platz 1,
D-07743 Jena, Germany}
\author{Radu Malureanu}
\affiliation{DTU Fotonik - Department of Photonics Engineering,
Technical University of Denmark, {\O}rsteds pl. 343, DK-2800 Kongens
Lyngby, Denmark}
\author{Falk Lederer}
\affiliation{Institute of Condensed Matter Theory and Solid State Optics,
Friedrich-Schiller-Universit\"at Jena, Max-Wien-Platz 1, D-07743 Jena, Germany}
\author{Andrei Lavrinenko}
\affiliation{DTU Fotonik - Department of Photonics Engineering,
Technical University of Denmark, {\O}rsteds pl. 343, DK-2800 Kongens
Lyngby, Denmark}

\date{\today}

\begin{abstract}
Homogenization of metamaterials is a crucial issue as it allows to describe their
optical response in terms of effective wave parameters as e.g. propagation
constants. In this paper we consider the possible homogenization of chiral
metamaterials. We show that for meta-atoms of a certain size a critical density
exists above which increasing coupling between neighboring meta-atoms
prevails a reasonable homogenization. On the contrary, a dilution in excess will
induce features reminiscent to photonic crystals likewise prevailing a
homogenization. Based on Bloch mode dispersion we introduce an analytical
criterion for performing the homogenization  and a tool to predict the homogenization
limit. We show that strong coupling between meta-atoms of chiral metamaterials may prevent their homogenization at all.
\end{abstract}

%\keywords{}

\maketitle

\section{Introduction}

At present chiral metamaterials (MMs) attract a great deal of interest of the
scientific community since their potential to strongly affect the polarization state
of light is promising for many applications. There are many manifestations of this
potential of chiral MMs as giant optical activity
\cite{decker_strong_2009,plum_giant_2007-1,rogacheva_giant_2006}, circular
dichroism \cite{decker_circular_2007,gansel_gold_2009,kwon_optical_2008}
and negative index of refraction
\cite{tretyakov_s._waves_2003,pendry_chiral_2004,zhou_negative_2009,wang_nonplanar_2009,zhang_negative_2009,Wiltshire2009},
which are either considerably enhanced in comparison with the naturally
occurring phenomena or cannot be observed in nature at all.

By definition, the unit cell of a chiral material is not superimposable with its mirror image
and any chiral material exhibits gyrotropy in general. However, a giant gyrotropy
requires strong intrinsic resonances of the elementary
constitutive components, i.e. the chiral "meta-atoms". For non-resonant
structures the optical rotary power per meta-atom layer is usually vanishingly
small. Even for resonant structures the achievable polarization rotation is often
below the requirements of applications. This limitation can be overcome
by stacking of several functional layers. Conventionally, light propagation in
multilayered MMs is described by assigning effective wave parameters (EPs) to this
MM \cite{Smith2002, menzel_retrieving_2008, plum_metamaterial_2009, wang_chiral_2009, zhang_negative_2009, zhou_negative_2009, xiong_construction_2010}, such as e.g. the propagation constant $k$ or the effective refractive index introduced as $n=k c/\omega$ . Such
EPs simplify the design and the description of functional MM devices. However,
before assigning EPs it is of utmost importance to assure that the respective MM
may be actually homogenized \cite{Simovski2009}.

When the mutual coupling between adjacent functional layers in a stack is
strong, the convergence of e.g. the effective refractive index with the stack
thickness towards its bulk value is poor\cite{Zhou2009}. Then, and this is a common situation,
the EPs depend on the MM slab thickness. Such effective properties have very
limited sense \cite{Simovski2007}, since they cannot predict the response (for
example, reflectance and transmittance) of a MM slab with an arbitrary
thickness. The EPs are very rarely identical for a single and an infinite number of
monolayers, though designs exist exhibiting this property
\cite{Andryieuski2009}. Near field coupling  has been investigated for the constitutive elements of
meta-atoms \cite{liu_three-dimensional_2008,sersic_electric_2009,aydin2010symmetry},
magnetic plasmon structures \cite{shamonina_properties_2006,liu_coupled_2009} and chiral MM
bi-layers \cite{li_coupling_2010}. The latter work [\onlinecite{li_coupling_2010}] showed that the functional properties of the bi-layered MM with coupling between layers are different from the properties of a monolayer. However, up to our knowledge, the general analysis of the coupling effects on the EPs of chiral metamaterials has not been conducted yet.

The main purpose of this work is to provide a systematic and comprehensive
analysis of the influence of the coupling effects on the EPs of chiral MMs and on the chiral MMs homogenization. We
introduce an analytical criterion of homogeneity based on the dispersion
relation of Bloch modes. We show that several chiral MMs can be homogenized and in such case there is an optimal MM period, or in other words, an
optimal distance between MM layers. We also show that in other cases MMs
cannot be homogenized in the spectral resonance domain despite of a
weakening of the coupling among adjacent layers by increasing their separation.

We consider only effective wave parameters
rather than any material EPs \cite{Simovski2007, Menzel2008}.  Hence, effects
caused by the mesoscopic structure of MMs (nonlocality, spatial dispersion, see
e.g. [\onlinecite{Simovski2007,Menzel2010}]) are not in the focus of
this contribution. The aim is rather to show under which conditions a
multilayered periodic chiral MM may be described as an effectively homogeneous medium, similar to the investigations performed for the fishnet MM \cite{Rockstuhl2008,yang_2010} and multilayered dielectric stack in \cite{Mortensen2010}.
To avoid any complications with the tensorial properties of
homogenized materials, the direction of propagation is fixed along the axis
pointing normal to the MM surface and the responses of the media are probed
upon illumination with left- and right-circular polarized waves only. Because of
normal incidence  the eigenvalue problem reduces to a scalar problem.

After having introduced the methodology of the research in Sec. II, we analyze
as an example the twisted-cross MM \cite{decker_strong_2009} with respect to
its homogenization in Sec. III. Additionally we introduce an analytical homogenization criterion. As a
counterexample for a structure that cannot be homogenized, the twisted
split-ring resonator MM \cite{decker_twisted_2010} is considered in Sec. IV. The importance of coupling effects for chiral MMs is emphasized in the Sec. V. Mechanical analogy of the periodic metamaterial is also provided there. The Conclusions section sums up the work.

\section{Methodology}

For an adequate description of a MM as homogeneous several requirements have to be fulfilled.

\begin{enumerate}
    \item The operating frequency must be below the first Bragg resonance $\Re(ka)<\pi$, where $a$ is the meta-atom size in the direction of wave propagation. Otherwise, the MM constitutes a photonic crystal and its homogenization is meaningless \cite{Simovski2009}.
    \item The lateral MM periodic must be such that the propagation of nonzero diffraction orders is suppressed \cite{Menzel2010}, i.e., the transverse size of a unit cell is required to be smaller than half the wavelength.
    \item For each polarization state the light propagation must be governed by a single (fundamental) Bloch mode only \cite{Menzel2010} having the smallest propagation losses.
    \item The incident light should predominantly couple to this Bloch mode \cite{Rockstuhl2008, Menzel2010}.
\end{enumerate}

In general, situations may occur where light couples to a higher-order Bloch mode
more efficiently than to the fundamental one. If this higher-order Bloch mode is
more damped than the fundamental one, its impact becomes negligible above a
certain thickness of the MM slab and EPs converge to those of the bulk.
However, if the attenuation coefficients for all Bloch modes to which light can
couple are comparable, then this leads to thickness-dependent EPs and
results in the lack of homogeneity.

As the tool to analyze such questions we rely on the dispersion relation of Bloch
modes. The respective dispersion diagrams provide comprehensive information
about the eigenmodes that are sustained by the MM. The occurrence of a Bragg
resonance can be extracted from the dispersion relation and occurs whenever
$\Re(ka)=\pi$ holds. From the analysis of the dispersion of the imaginary part of
the wavenumber, $\Im(k)$, we can evaluate the propagation losses. So it is easy
to infer about requirements (iii) and (iv) - number of dominating modes and their
losses by considering in detail the dispersion relation. Thus, the Bloch modes
dispersion diagram is an efficient mean of MM homogeneity characterization.
We would like to remark that in the case of chiral media with $C_4$-symmetry there are two
fundamental modes: right and left circular polarized (RCP and LCP).

The chiral properties of a MM are determined through their effective refractive
indices for the LCP and RCP waves the $n_{L}$ and $n_{R}$,
respectively. The chirality parameter is given by $\kappa=(n_R-n_L)/2$. Its real
part $\Re(\kappa)$ describes the rotation of the polarization ellipse (or optical
activity) \cite{menzel_retrieving_2008} whereas the imaginary part
$\Im(\kappa)$ governs the circular dichroism.

\begin{figure}
\includegraphics{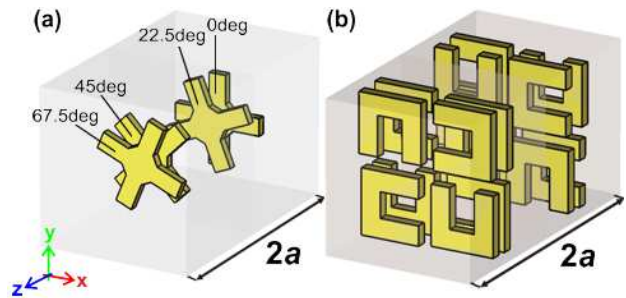}%
\caption{\label{Designs}(Color online). Metamaterial designs under
consideration: TC (a) and TSRR
(b). Wave propagation and metamaterial stacking direction is along
$z$-axis.}
\end{figure}

The first homogeneity requirement implies that a smaller lattice period $a$ in the
propagation direction is beneficial. To check this statement we calculate the
Bloch modes dispersion diagrams for both the twisted-cross metamaterials
\cite{decker_strong_2009} (Fig.~\ref{Designs}a) and twisted split-ring resonator
metamaterials \cite{decker_twisted_2010} (Fig.~\ref{Designs}b). The lateral sizes of their unit cells are such that the homogeneity requirement (ii) is satisfied in the
whole frequency range of interest. These examples were only chosen for the
sake of proving the applicability of our approach and constitute referential unit
cells that are currently under investigation.

To simulate the optical response of MM slabs with various thicknesses we rely
on the commercially available CST Microwave Studio software \cite{CST} that
implements the finite-integral method in the frequency domain. The thin slab EPs are
retrieved from the complex reflection/transmission coefficients
\cite{plum_metamaterial_2009}, which we refer to as the standard method. The thick slab EPs, which are close to bulk parameters, are restored from
the spatial dependence of the wave field with the wave propagation retrieval
method \cite{Andryieuski2009a, Andryieuski2010}. The Bloch dispersion
diagram calculations are performed for infinite periodic media using a plane
wave expansion method \cite{li2003} and taking into account the dispersion of
the metal permittivity. Details of the calculation method can be found elsewhere
\cite{Rockstuhl2008a, Rockstuhl2008, Menzel2010}.

Comparing the spectral simulations of finite structures performed with the CST
Microwave Studio with those done with the plane
wave expansion method we observed a steady blue shift in spectra of about 20
THz for the latter method. This shift that happens due to different numerical approaches is of minor importance for our conclusions.

\section{Twisted cross metamaterial}

The twisted-cross (TC) MM \cite{decker_strong_2009} (see Fig.~\ref{Designs}a)
is an example of a chiral MM that provides strong optical activity. Its
meta-atoms consist of two golden crosses embedded into a dielectric. One
cross is 22.5 degrees twisted with respect to  the other. Lateral unit cell sizes are $a_x=a_y=500$ nm. The metallic crosses of thickness 25 nm are separated with a dielectric spacer of 37.5 nm. The eigenfunctions of
the TC are RCP and LCP because of the $C_4$-symmetry of the MM. For our
investigation we use the parameters of Ref. [\onlinecite{decker_strong_2009}].
However, in Ref. [\onlinecite{decker_strong_2009}] only a monolayer is
characterized. To access the bulk properties of such a MM we stack multiple
monolayers as presented in Fig.~\ref{Designs}a. The meta-atom size in the propagation
direction $a$ is subject to variations. The crosses in each meta-atom are twisted by 45 degrees with respect to
each other. Details can be seen in Fig.~\ref{Designs}a where the angles of
consecutive layers are indicated.

Note that, although the real unit cell size in propagation direction is $2a$,
the size $a$ is actually the length scale of importance because it denotes the periodic
distance of consecutive functional layers. Since the optical response of the two
parts of the unit cell are close to identical, we wish to regard this distance
$a$ as the period. In fact, calculating the dispersion diagrams for the structure
with the second double cross being identical to the first one, i.e. leading to a
true period $a$, results qualitatively in identical characteristics. This holds
except for the case of dense stacking, where the material becomes achiral. The
Bragg resonances for the present structure are caused by the reflection of light
at the consecutive layers of the unit cell rather than by the boundary between
two unit cells. Bragg gaps due to the period $2a$ are not observed. Therefore,
we will call the parameter $a$ in the following the period of the structure.

\begin{figure}
\includegraphics{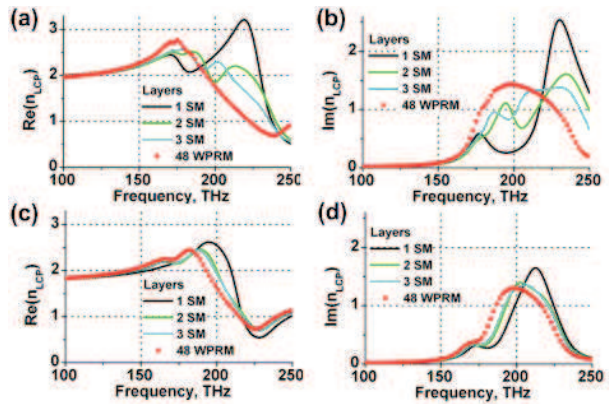}%
\caption{\label{TCEPs}(Color online). Effective refractive index of
left circular polarized light of a twisted-cross metamaterial as a
function of frequency and for different numbers of layers that form the MM.
Dense stacking (period $2a=250$ nm) $\Re(n_{LCP})$ (a), $\Im(n_{LCP})$ (b);
sparse stacking (period $2a=500$ nm) $\Re(n_{LCP})$ (c), $\Im(n_{LCP})$
(d). Effective parameters are retrieved with the standard
method\cite{plum_metamaterial_2009} (SM) for one (black line), two
(green/grey line) and three (cyan/light grey line) monolayers and
with the wave propagation retrieval method \cite{Andryieuski2010}
(WPRM) for 48 layers (red/dark grey dots).}
\end{figure}

To challenge the third requirement of the homogeneity we applied the standard
retrieving (based on reflection/transmission) to
thin MM slabs and the wave propagation method to
thick MM slabs. The aim of such a retrieval with different methods was to check
the convergence of the effective parameters. We studied MMs with two
different center-to-center separations, the dense ($a=125$ nm) and the sparse
stacking ($a=250$ nm). The thin-slab case consisted of 1-3 meta-atom layers
for both MMs. For the thick slab EPs retrieval, 48 monolayers were stacked,
thus mimicking the bulk material.

The structures exhibit completely different types of convergence with the slab
thickness. For the dense stacking the convergence in the resonant region is very
slow (Fig.~\ref{TCEPs}a and b). So we suspect that the TC is not homogeneous
there due to strong coupling between consecutive unit cells. The coupling can be
reduced by separating the unit cells more. Indeed, for
the sparse stacking, the effective refractive index of the thin slab quickly converges to
the bulk values that are retrieved with the wave propagation retrieval method (Fig.~\ref{TCEPs}c and d). This
is a clear indication that such MMs might satisfy the homogeneity conditions.

\begin{figure}
\includegraphics{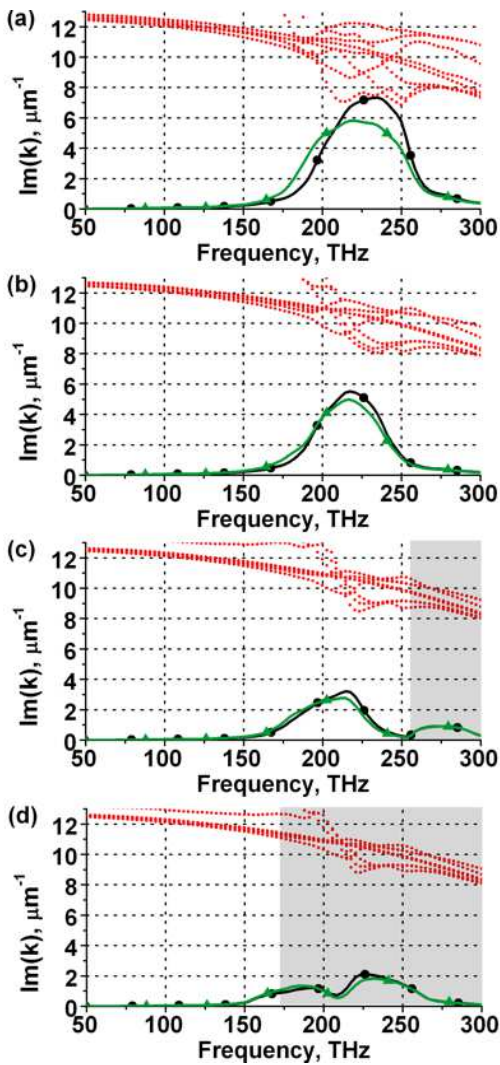}
\caption{\label{TCMMDispersion}(Color online). Bloch modes damping $\Im(k)$
dispersion spectra for twisted-cross MM for different separations:
 125 nm (a), 250 nm (b), 400 nm (c), 500 nm (d). Fundamental
modes are shown as black line with circles and green/grey line with
triangles, all the higher-order Bloch modes are shown in red/dark
grey dots. The frequency region above the first Bragg resonance is
shadowed.}
\end{figure}

To check the limits where TC homogenization fails we calculated the Bloch modes
dispersion diagram for increasing separation, as shown in Fig.~\ref{TCMMDispersion}. It is clearly seen that for the
dense stacking ($a=125$ nm) higher-order Bloch modes have a damping comparable
with the fundamental modes in the resonant region
(Fig.~\ref{TCMMDispersion}a). Slow convergence of the EPs as seen in
Fig.~\ref{TCEPs}a, b, confirms that light couples to the higher-order Bloch
modes, thus homogeneity requirement (iv) is violated.

For the sparse stacking case of $a=250$ nm (Fig.~\ref{TCMMDispersion}b) the
continuum of the higher-order Bloch modes is separated from the fundamental
ones in terms of losses. The higher-order modes are  much stronger damped
than the fundamental ones. Thus one may expect that only the fundamental
modes govern the light propagation. Moreover, the MM unit cells are still
sub-wavelength (the resonant response occurs at frequencies far below the
lowest order Bragg resonance), so all four requirements are satisfied, and the TC
MM can be considered as a homogeneous.

Increasing the period to $a=400$ nm we observe a new peak in
$\Im(k)$ emerging at around $275$ THz and which is separated from the
lowest resonance by a local minimum in the losses at $250$ THz
(Fig.~\ref{TCMMDispersion}c). The Bragg condition $\Re(ka)=\pi$ (the plot for
$\Re(ka)$ not shown here) is satisfied between $250$ THz and $300$ THz,
therefore, this peak can be clearly identified as the first Bragg resonance. In the frequency
region above the first Bragg resonance ($\Re(ka)>\pi$) the
properties of periodic arrangement of unit cells dominate and, according to
requirement (i), homogenization fails. The frequency ranges where photonic
crystal properties prevail are shadowed in Fig.~\ref{TCMMDispersion}.

Increasing $a$ even further to $500$ nm we face the situation where the first
Bragg resonance overlaps  with the intrinsic resonance of the meta-atoms at
$220$ THz (Fig.~\ref{TCMMDispersion}d). Interestingly, the dip in the absorption spectrum at 210 THz
appears instead of the peaks for all other separations
(Fig.~\ref{TCMMDispersion}a, b and c). This resembles the phenomenon of
electromagnetically induced transparency \cite{Boller1991, Liu2010}, when two
coupled oscillators with a small detuning show an absorption dip at the
frequency, where the individual oscillators resonance is observed.

If we compare the magnitudes of mode losses extracted from
Fig.~\ref{TCMMDispersion}, we conclude that losses of the higher-order Bloch
modes remain almost constant while decreasing the unit cell size. On the
contrary, the losses of the fundamental modes increase considerably for smaller
separation due to the increase of the metal filling fraction.

\begin{figure}
\includegraphics{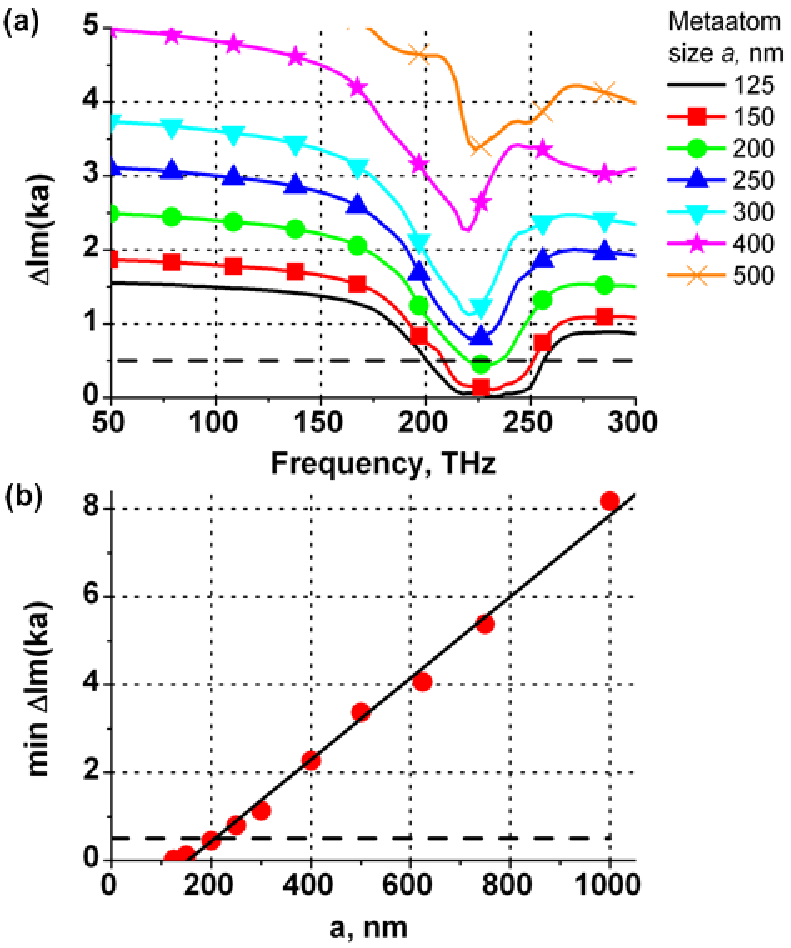}%
\caption{\label{TCMMDeltaka}(Color online). (a) The difference of the
propagation constants
$\Delta\Im(ka)=\min_{i\geq3}[\Im(k_{i}a)]-\max[\Im(k_1a,k_2a)]$
of the least damped higher-order Bloch mode and the most damped
fundamental mode for various unit cell sizes $a$: 125 nm (black
line), 150 nm (red line with squares), 200 nm (green line with
circles), 250 nm (blue line with triangles up), 300 nm (orange line
with triangles down), 400 nm (magenta line with stars), 500 nm
(brown line with crosses). (b) The minimum of $\min \Delta\Im(ka)$
in (a) as a function of the period  (red
circles). The straight line represents a linear fit of $\min
\Delta\Im(ka)$ on $a$. The dashed line corresponds to the homogenization limit
$\Delta\Im(ka)=0.5$. }
\end{figure}

The brief summary of the presented results is that there exists a range of
optimal longitudinal  periods to achieve the best homogeneity for the
twisted-cross MM, which is limited by the influence of the higher-order Bloch modes from one side and the Bragg resonance from the other side. To determine it we plot the difference $\Delta
\Im(ka)=\min_{i\geq3}(\Im(k_{i}a))-\max(\Im(k_1a,k_2a))$ between the
attenuation constants of the least damped higher-order Bloch mode and the
most damped fundamental mode  (Fig.~ \ref{TCMMDeltaka}a). We suggest to
neglect higher-order Bloch modes in the consideration if the difference in
damping per unit cell is $\Delta \Im(ka)\geq0.5$. In this case the intensity of the
higher-order modes will be attenuated $e$ times more over the period
$a$, compared to the fundamental modes. In the resonant region between 200
and 250 THz the homogeneity condition $\Delta\Im(ka)>0.5$ is satisfied for
$a\geq200$ nm (Fig.~\ref{TCMMDeltaka}a). Outside the resonant
region this condition is much more moderate and the TC can be homogenized for
smaller values of $a$.

The dependence of the minimum of the damping difference $\min
\Delta\Im(ka)$ as a function of the period $a$ is shown in
Fig.~\ref{TCMMDeltaka}b. Surprisingly, this dependence is not only
monotonously rising but also close to linear. This fact points out the predictive
potential of the used methodology. Thus, it suffices to calculate the Bloch modes
for only two periods in order to approximately predict the homogenization limit
for the designed metamaterial.

We also checked the homogenization criterion $\Delta\Im(ka)>0.5$ for
the split cube in cage negative index MM \cite{Andryieuski2009}. This MM shows very fast convergence of the EPs with the number of monolayers - the refractive index is the same from one monolayer to the infinite number of monolayers. In the whole
frequency range of interest higher-order Bloch modes are well separated from
the fundamental ones $(\Delta\Im(ka)>1.5)$ that explains why the EPs of the split-cube in cage
do not depend on the slab thickness.

\section{Twisted split-ring resonator metamaterial}

It is natural to make a hypothesis that for some metamaterial designs with strong coupling between
consecutive layers, the MM cannot be homogenized  in the resonance region at
all. In other words,  $\Delta\Im(ka)< 0.5$ even for the large separation of
meta-atoms where homogenization is prevented by the appearance of Bragg
resonances. To check this hypothesis we investigated the twisted split-ring resonator
(TSRR) MM \cite{decker_twisted_2010} (see Fig.~\ref{Designs}b). It consists of the golden U-shaped split-ring resonators embedded in a dielectric. The in-plane lattice constants are $a_x=a_y=885$ nm. The split rings of thickness 60 nm are separated by a dielectric spacer of 85 nm. All the geometrical and material
parameters are taken as in Ref.[\onlinecite{decker_twisted_2010}] except the
period $a$.

\begin{figure}
\includegraphics{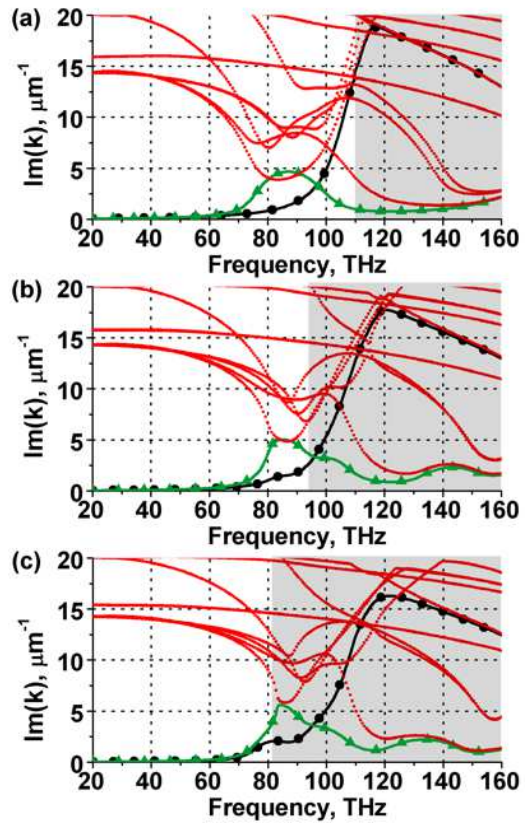}
\caption{\label{TSRRMMdispersion}(Color online). The loss spectrum
 for TSRR MM for different periods: 290 nm
(a), 450 nm (b), 600 nm (c). Fundamental modes are shown as black
lines with dots and green/grey lines with triangles, all
higher-order Bloch modes are shown as red/dark grey dots. The
frequency region above the first Bragg resonance is shadowed.}
\end{figure}

We calculated the Bloch modes dispersion diagram for three center-to-center
separations of the monolayers: $a=290$ nm, $a=450$ nm and $a=600$ nm.
One of the fundamental modes designated by the black solid line in
Fig.~\ref{TSRRMMdispersion} fails to satisfy the homogeneity requirements even
for the sparsest stacking at $a=600$ nm. Its losses are higher than those for a
few of the higher-order modes. The other fundamental mode
(Fig.~\ref{TSRRMMdispersion} green line) exhibits a comparable $\Im(ka)$ with
one of the higher-order Bloch mode for $a=290$ nm and $a=450$ nm. For the
sparsest stacking $a=600$ nm, this fundamental mode experiences a slightly
lower, but still unsufficient for homogenization, damping, since $\Delta(\Im(ka)) = 0.09 < 0.5$. A further increase of
the lattice period to $a=600$ nm shifts the real part of the wavevector
into the first band gap ($\Re(ka)=\pi$) for frequencies above 80 THz, so the
requirement (i) is not met. We would hardly term a metamaterial
homogeneous if only one eigenwave (for example, RCP) fulfills the
homogenization requirements. We may conclude that the TSRR does not
satisfy requirement (iii) around the resonance frequency for any period.
This is caused by the strong coupling between its meta-atoms. Thus, we prove our hypothesis that there exist MMs that cannot be considered homogeneous at the resonance frequency region.

It is important to mention that such MM keeps its optical functionality, i.e.
affecting the state of polarization. However, one must not extrapolate the
properties of a single or a few functional layers towards a stack of more
functional layers.

\section{Discussion}

The homogeneity criterion (i) requires a unit cell/wavelength ratio to be as small as
possible. However, in this case, as shown above,  in the resonance region
requirements (iii) (single Bloch mode)  and (iv) (efficient coupling of light to this
specific mode)  are frequently not met. When decreasing the unit cell size $a$, the fundamental modes' damping
increases (see Fig.~\ref{TCMMDispersion}), so sooner or later it becomes
comparable with the higher-order Bloch modes' damping. The higher-order Bloch modes then have a
comparable impact on the wave propagation in a MM slab as a direct
consequence of the strong interaction of the meta-atom monolayers.

\begin{figure}
\includegraphics{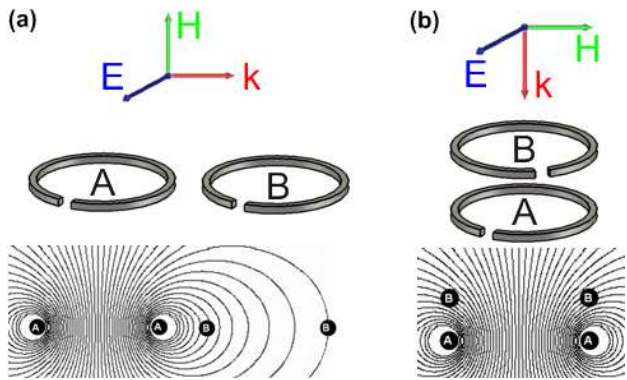}%
\caption{\label{Loops} The sketch of the split-ring resonators stacking in the
negative-permeability configuration (a) and in the chiral
configuration (b). Arrows represent the direction of the wave propagation, electric and magnetic field vectors. Black lines in the lower pictures represent the cross-section view of the lines
of the magnetic field created by SR A and penetrating SR B.}
\end{figure}

This conclusion, which
is basically applicable to any MM, is extremely important for chiral MMs where
the coupling of monolayers is usually strong. For example, for a
negative-permeability configuration of the split-ring (SR) resonator metamaterial in
the side-coupled arrangement, where the wavevector lies in the SR plane, SRs
are weakly coupled due to the mutual inductance (magnetic field penetration)
(see the sketch in Fig.~\ref{Loops}a). The induced current in  SR A creates a magnetic field that
penetrates SR B. However, the number of the field lines penetrating SR B is not
very large. On the contrary, for a chiral configuration, when the stacking
direction is perpendicular to the SR plane (see Fig.~\ref{Loops}b), the magnetic
field of SR A strongly penetrates SR B.

Most chiral metamaterials consist of split-ring resonators or gammadions,
where the meta-atoms are stacked similarly as in Fig.~\ref{Loops}b. The other
configuration, where the magnetic field of one meta-atom penetrates weakly
into the other one, corresponds to the twisted-cross MM. It is basically an
extension of the cut-wire pairs towards $C_4$-symmetry. The currents excited
in the double crosses create the magnetic field perpendicular to the stacking
direction and are thus weakly penetrating into the adjacent monolayer (similar
as in Fig.~\ref{Loops}a).

\begin{figure}
\includegraphics{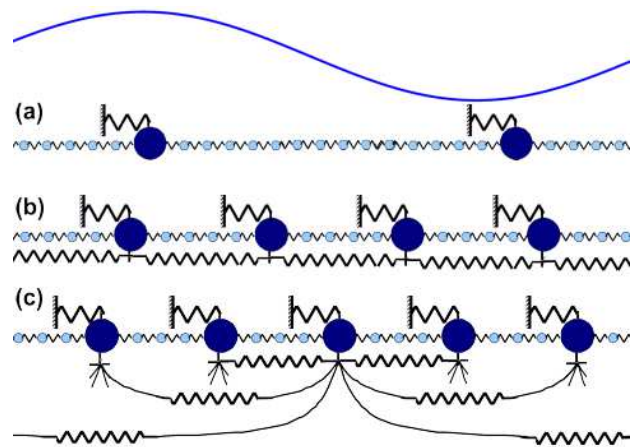}%
\caption{\label{Springball} Mechanical analogy of the metamaterial
in the photonic crystal regime (a), homogeneous (b) and
inhomogeneous (c) medium. Small balls and springs represent host
medium. Large balls represent resonant inclusions.}
\end{figure}

Historically, e.g. by Newton, Bernoulli, Lagrange and others, wave phenomena
were frequently illustrated by taking advantage of mechanical analogies. The
historical overview of a periodic medium description with a spring-ball system can be found in
Brillouin's book \cite{brillouin2003wave}. We likewise employ a mechanical
spring-ball system to illustrate the coupling effect in our MM system, see
Fig.~\ref{Springball}.

The host dielectric medium is represented by chains of small balls and springs.
Meta-atoms are represented in Fig.~\ref{Springball} as massive oscillators
having their own resonance frequency.  When the distance between the
meta-atoms is large, so that $\Re(ka)>\pi$, the periodic medium is in the
photonic crystal regime (Fig.~\ref{Springball}a). The massive oscillators
periodically arranged in the system scatter the wave individually. Then
constructive or destructive interference leads to the formation of stop-bands and
pass-bands. When the distance between meta-atoms decreases, but coupling
between monolayers is still weak, we get a kind of coupled resonator
waveguide, where only nearest-neighbor coupling matters (Fig.~\ref{Springball}b). Such a waveguide has
one fundamental mode (two in case of chiral metamaterials), which is
dominating the light propagation. Thus the periodic medium may be
homogenized, and the extraordinary properties of
MMs originate from the meta-atom resonances. When the distance between the
meta-atoms is further reduced, the coupling gets stronger and the nearest-neighbor
scenario ceases to hold. Now higher-order modes come into play
(Fig.~\ref{Springball}c) and  participate in the energy transfer together with the
fundamental mode. So, this is the regime, when higher-order Bloch modes can
propagate in the structure, and homogenization is not feasible.

As can be inferred from the spring-ball model the effect of higher-order modes is
negligible if there is no strong interaction between the meta-atoms. In this case the
effective properties are solely determined  by the resonances of the individual
meta-atoms. However, to ensure a decoupling while the period has to remain
small in comparison to the wavelength, i.e., avoiding photonic crystal effects,
the constitutive elements of the meta-atoms have to  be very small. Obviously, in this case the resonance
features of the meta-atoms disappear for a fixed frequency. Such MM would
hardly exhibit any extraordinary properties.

\section{Conclusions}

To sum up, it turns out that the possibility of homogenization of chiral MM depends crucially on the coupling strength between meta-atoms. For weak coupling
between monolayers the MM can be homogenized. Then the optimal meta-atom
separation can be identified by looking at the Bloch mode dispersion relation in the frequency domain below
the Bragg resonance. The TC MM homogenization works best if the characteristic parameter $a$ of the unit cell is as small as
possible but the higher-order Bloch modes remain well separated from the
fundamental modes. In this contribution we propose the analytical homogenization criterion as $\Delta\Im(ka)>0.5$. In this respect it is helpful that the calculation of the
dispersion relation for two unit cell parameters $a$ suffices to estimate the lower homogeneity limit of a metamaterial.

For strong coupling between meta-atoms, the homogenization of MMs
might be impossible at all. In this case the functionality of a MM, e.g. a stack of several monolayers, must be characterized rather than retrieving EPs of one monolayer and
extrapolating them to the properties of a multilayered stack.

\begin{acknowledgments}
The authors wish to thank M.Wubs for critical comments on the manuscript. A.A., R.M. and A.L. acknowledge financial support from the Danish
Research Council for Technology and Production Sciences via the
NIMbus project and by COST Action MP0702. C.M., C.R. and F.L. acknowledge financial support
by the German Federal Ministry of Education and Research (Metamat,
PhoNa) and by the Thuringian State Government (MeMa).
\end{acknowledgments}

\bibliography{Andryieuski}

\end{document}